\documentclass[fleqn,twocolumn,superscriptaddress,aps]{revtex4}
\usepackage{epsfig}
\usepackage{graphicx}
\usepackage{color} 
\usepackage{amsfonts}
\usepackage{amsmath}

\newcommand{\bi}{\bibitem}
\newcommand{\be}{\begin{equation}}
\newcommand{\ee}{\end{equation}}
\newcommand{\ba}{\begin{eqnarray}}
\newcommand{\ea}{\end{eqnarray}}
\newcommand{\non}{\nonumber}
\newcommand{\bra}{\langle }
\newcommand{\ket}{\rangle}

\begin{document}
\title{Classical and quantum shortcuts to adiabaticity in a tilted piston}
\author{Ayoti Patra}
\affiliation{Department of Physics, University of Maryland, College Park, Maryland 20742, USA}
\author{Christopher Jarzynski}
\affiliation{Department of Chemistry and Biochemistry and Institute for Physical 
Science and Technology, University of Maryland, College Park, Maryland 20742, USA}

\begin{abstract}
Adiabatic quantum state evolution can be accelerated through a variety of {\it shortcuts to adiabaticity}.
In one approach, a counterdiabatic quantum Hamiltonian $\hat H_{CD}$ is constructed to suppress nonadiabatic excitations.
In the analogous classical problem, a counterdiabatic classical Hamiltonian $H_{CD}$ ensures that the classical action remains constant even under rapid driving.
Both the quantum and classical versions of this problem have been solved for the special case of scale-invariant driving, characterized by linear expansions, contractions or translations of the system.
Here we investigate an example of a non-scale-invariant system -- a {\it tilted piston}.
We solve exactly for the classical counterdiabatic Hamiltonian $H_{CD}(q,p,t)$, which we then quantize to obtain a Hermitian operator $\hat H_{CD}(t)$.
Using numerical simulations, we find that $\hat H_{CD}$ effectively suppresses non-adiabatic excitations under rapid driving.
These results offer a proof of principle -- beyond the special case of scale-invariant driving -- that quantum shortcuts to adiabaticity can successfully be constructed from their classical counterparts.
\end{abstract}

\maketitle

\section{Introduction}
According to the quantum adiabatic theorem \cite{Born28}, a wavefunction that is initialized in an eigenstate of a time-dependent Hamiltonian will subsequently follow the instantaneous energy eigenstate, 
provided the Hamiltonian is driven  sufficiently slowly. While maintaining a system in its instantaneous energy eigenstate is essential in problems involving 
coherent quantum control
\cite{Avron88,Kral07}, slow driving may render the system vulnerable to decoherence, thereby compromising the fidelity of the desired evolution. {\it Shortcuts to 
adiabaticity} (STA) are strategies for achieving adiabatic results with fast driving protocols \cite{Torrontegui13}. STA have been studied theoretically in contexts 
such as adiabatic quantum computation \cite{Nielsen}, finite time thermodynamics \cite{Andresen84}, quantum sensing and metrology \cite{Giovanetti06} and quantum 
simulation \cite{Trabesinger12,Bason11}. Experimental studies have been carried out using cold atom systems \cite{Schaff10,Schaff11}, trapped ions \cite{Walther12,
Bowler12} and two-level systems \cite{Zhang13} subjected to fast driving.
 
Methods for constructing STA include inverse engineering using Lewis-Riesenfeld invariants \cite{Chen10,Chen11}, inversion of scaling laws 
\cite{dCampo11,dCampo12} and fast-forward techniques \cite{Masuda09,Masuda11,Torrontegui12}. Of particular relevance for the present paper is {\it transitionless quantum driving} 
(TQD), proposed independently by Demirplak and Rice \cite{Rice03,Rice05}, and Berry \cite{Berry09}, and described briefly in the following paragraph.
  
Consider a Hamiltonian ${\hat H}_0(\boldsymbol{\lambda})$, where $\boldsymbol{\lambda}$ denotes a vector of externally controlled parameters that are varied with time according to a protocol  $\boldsymbol{\lambda}(t)$. In the TQD approach, a counterdiabatic (CD) Hamiltonian ${\hat H}_{CD}(t)$ is constructed, such that when the system evolves under
\be
{\hat H}(t) = {\hat H}_0(\boldsymbol{\lambda}(t))+{\hat H}_{CD}(t), 
\label{fullHamil}
\ee
it follows the adiabatic trajectory of ${\hat H}_0(\boldsymbol{\lambda}(t))$ even for rapid driving, i.e.\ the term ${\hat H}_{CD}(t)$ suppresses non-adiabatic excitations.
The exact expression for the CD Hamiltonian can be written as  \cite{Berry09}
 \ba
   {\hat H}_{CD}(t)  &=& i \hbar {\boldsymbol{\dot \lambda}}\sum\limits_{m} \Big(|\partial_{\boldsymbol{\lambda}}
  m \ket \bra m| - \bra m |\partial_{\boldsymbol{\lambda}} m \ket  |m \ket \bra m| \Big)\non \\
  &\equiv& {\boldsymbol{\dot \lambda}} \cdot \hat{\boldsymbol{\xi}}(\boldsymbol{\lambda}(t)).
  \label{Hqlambda}
 \ea
The sum is taken over eigenstates $|m (\boldsymbol {\lambda}) \ket$ of ${\hat H}_0(\boldsymbol{\lambda})$, and 
$|\partial_{\boldsymbol{\lambda}} m  \ket \equiv  \partial_{\boldsymbol{\lambda}} |m (\boldsymbol {\lambda}) \ket$. 
The operator $\hat{\boldsymbol{\xi}}(\boldsymbol{\lambda})$ acts as a generator of adiabatic evolution \cite{Jarzynski13}.
The more rapidly the parameters are varied, the greater the magnitude of the term ${\hat H}_{CD}={\boldsymbol{\dot \lambda}} \cdot \hat{\boldsymbol{\xi}}$ needed to suppress excitations \cite{Demirplak08,dCampo12p}.
 
Eq.(\ref{Hqlambda}) is exact, but to gain physical insight 
 and to pave the way for experimental implementation it is desirable to rewrite ${\hat H}_{CD}(t)$ in terms of operators corresponding to physical observables 
 such as position and momentum. For a harmonic oscillator with a time dependent stiffness $k(t)$, Eq.(\ref{Hqlambda}) reduces 
 to:  \cite{Chen11}
 \be 
 {\hat H}_{CD} = -\frac{{\dot k}}{8k}({\hat p}{\hat q} + {\hat q} {\hat p}).
 \label{SHOpq}
 \ee
Similar expressions hold for a particle in a box \cite{dCampo12}, for attractive power law potentials \cite{Jarzynski13}, and more generally for arbitrary potentials undergoing \textit{scale invariant driving} (see Eq.(\ref{eq:scaleInvt})), characterized by simple expansion, contraction or translation of the potential \cite{dCampo13,Deffner14}.
However, for general ${\hat H}_0(\boldsymbol{\lambda}(t))$ it is not clear how to rewrite Eq.(\ref{Hqlambda}) in terms of operators such as ${\hat q}$ and ${\hat p}$. 

In Ref.~\cite{Jarzynski13} it was proposed that this problem can usefully be approached by studying the classical counterpart of TQD: {\it dissipationless classical driving} (DCD).
In one degree of freedom, the 
classical adiabatic invariant is the action $I_0(q,p;\boldsymbol{\lambda}) = \oint p'\,{\rm d}q'$, equivalently the volume of phase space enclosed by a surface of 
constant energy \cite{Goldstein}. The  problem of DCD is formulated as follows: For a time-dependent Hamiltonian $H_0(q,p;{\boldsymbol \lambda}(t))$, find the
counterdiabatic term $H_{CD}(q,p,t) =  \boldsymbol{{\dot \lambda}} \cdot \boldsymbol{\xi}(q,p;\boldsymbol{\lambda}(t))$ 
such that the action $I_0(q,p;\boldsymbol{\lambda}(t))$ (defined with respect to $H_0$) remains constant along any trajectory evolving under the Hamiltonian 
\be
H(t) =  H_0(\boldsymbol{\lambda}(t))+{\boldsymbol{\dot \lambda}} \cdot \boldsymbol{\xi}(\boldsymbol{\lambda}(t)).
\label{fullHamilC}
\ee

A natural question then arises: if we solve for the classical generator $\boldsymbol{\xi}(q,p;\boldsymbol{\lambda})$ and then quantize it to obtain an operator $\hat{\boldsymbol{\xi}}(\boldsymbol{\lambda})$, will the term ${\boldsymbol{\dot \lambda}} \cdot \hat{\boldsymbol{\xi}}(\boldsymbol{\lambda}(t))$ suppress non-adiabatic transitions under quantum evolution?
In other words, can we construct ${\hat H}_{CD}$ -- either exactly or approximately -- by first obtaining its classical counterpart and then quantizing it?

This strategy for constructing ${\hat H}_{CD}$ was proposed in Ref.\cite{Jarzynski13}, where it was shown that it yields the correct result for attractive power law potentials, including the harmonic oscillator and the particle in a box as limiting cases.
This encouraging result was generalized to arbitrary potentials undergoing scale-invariant driving (Eq.(\ref{eq:scaleInvt})) \cite{dCampo13,Deffner14}.
In all these cases the classical counterdiabatic term takes the form $H_{CD}=g(t) p + h(t) qp$, and its quantized counterpart
\be
\label{HCD_scaleInvt}
\hat H_{CD} = g(t) \hat p + \frac{h(t)}{2} \left( {\hat p}{\hat q} + {\hat q} {\hat p} \right)
\ee
can be shown to be equivalent to Eq.(\ref{Hqlambda}).

 In this work, we investigate whether this strategy succeeds for {\it non}-scale-invariant driving protocols.
We study a test case: a particle in a box with infinite 
 walls and a slanted base, i.e. a \textit{tilted piston}. In Sec. II, we briefly review dissipationless classical driving, and we specify the Hamiltonian and 
 the driving protocols that will be studied. We solve exactly for $H_{CD}(q,p,t)$ in Sec. III, and we quantize it semiclassically in Sec. IV.
 Finally we study numerically whether the resulting quantum operator $\hat{H}_{CD}(t)$ produces the desired transitionless quantum 
 driving. Details of the numerical approach are described in Sec. V, and the results are presented in Sec. VI. We conclude briefly in Sec. VII.

 \section{Classical Dissipationless driving and the system under study} 
 \label{sec:DCD}
 
 Let $z=(p,q)$ denote a point in the system's two-dimensional phase space.
 The counterdiabatic Hamiltonian $H_{CD}(z;t) =  \boldsymbol{\lambda} \cdot \boldsymbol{\xi}(z;\boldsymbol{\lambda}(t))$, when added to the unperturbed Hamiltonian 
 $H_0(z;\boldsymbol{\lambda}(t))$ (see Eq.(\ref{fullHamil})), ensures that the classical system follows an adiabatic trajectory -- along which the action $I_0$ is constant
 -- even when the driving is rapid.
Below we briefly summarize how $\boldsymbol{\xi}(z;\boldsymbol{\lambda})$ is constructed, as described in greater detail in Ref \cite{Jarzynski13}.
 
 Let the microcanonical average of a quantity be denoted by
 \be
\bra \dots \ket_{E,\boldsymbol{\lambda}} \equiv \frac{1}{\partial_E \Omega} \int dz \delta(E-H_0) \dots,
\label{MCdef}
\ee
and the volume of phase space enclosed by an energy shell $E$ be denoted by  
\be
\Omega(E,\boldsymbol{\lambda}) \equiv \int dz \theta \left[E-H_0(z;\boldsymbol{\lambda}) \right].
\label{psvol}
\ee
Then the desired classical generator $\boldsymbol{\xi}$ satisfies \cite{Jarzynski13}: 
\be
\boldsymbol{\xi}(z_b;\boldsymbol{\lambda})-\boldsymbol{\xi}(z_b;\boldsymbol{\lambda})=\int_a^b dt \boldsymbol{\nabla} \tilde{H}_0(z(t);\boldsymbol{\lambda}),
\label{xiexp}
\ee
where $z_a$ and $z_b$ are two points on the energy shell $E$, $z(t)$ is a trajectory that evolves from $z_a$ to $z_b$ under
$H_0(z; \boldsymbol{\lambda})$ (with $\boldsymbol{\lambda}$ fixed), and the integrand is defined as
$\boldsymbol{\nabla}\tilde{H}_0 \equiv \boldsymbol{\nabla} H_0 - \bra \boldsymbol{\nabla} H_0 \ket_{E,\boldsymbol{\lambda}}$, with 
$\boldsymbol{\nabla} \equiv \partial/\partial\boldsymbol{\lambda}$.
By convention the microcanonical average of $\boldsymbol{\xi}$ is set to zero, 
\be
\bra \boldsymbol{\xi} \ket_{E, \boldsymbol{\lambda}} =0.
\label{xiavg}
\ee
Eqs.(\ref{xiexp}) and (\ref{xiavg}) uniquely specify the generator $\boldsymbol{\xi}(z;\boldsymbol{\lambda})$. 

For a  system with one degree of freedom, the time average and the microcanonical average of a quantity are equivalent, therefore we can compute 
$\bra \boldsymbol{\nabla} H_0 \ket_{E,\boldsymbol{\lambda}}$ by evaluating the time average of $\boldsymbol{\nabla} H_0$ along a periodic trajectory of energy $E$.
Alternatively, this microcanonical average can be determined by defining the inverse function 
$E(\Omega,\boldsymbol{\lambda})$ from $\Omega(E,\boldsymbol{\lambda})$, and by using the cyclic identity of partial derivatives:
\be
\boldsymbol{\nabla} E(\Omega,\boldsymbol{\lambda}) = -\frac{\boldsymbol{\nabla} \Omega(E,\boldsymbol{\lambda})}{\partial_E \Omega(E,
\boldsymbol{\lambda})} = \bra \boldsymbol{\nabla} H_0 \ket_{E,\boldsymbol{\lambda}}.
\label{avgH}
\ee

For a harmonic oscillator with a time dependent stiffness $k(t)$, the procedure described above leads to the classical counterdiabatic Hamiltonian
 \be 
H_{CD}(z,t) = -\frac{\dot k}{4k}qp.
 \ee
Upon quantization, this result agrees with the quantum counterdiabatic Hamiltonian for the harmonic oscillator, Eq.(\ref{SHOpq}), which was originally obtained by direct 
evaluation of the Demirplak-Rice-Berry formula, Eq.(\ref{Hqlambda}).
More generally, a time-dependent potential of the form
\be
\label{eq:scaleInvt}
V(q;f,\gamma) = \frac{1}{\gamma^2} V_0\left(\frac{q-f}{\gamma}\right)
\ee
where $f=f(t)$ and $\gamma=\gamma(t)$, is said to undergo {\it scale-invariant driving}.
For scale-invariant driving, Eqs.(\ref{xiexp}) and (\ref{xiavg}) lead to a simple expression for $H_{CD}(z;t)$ that, upon quantization, give the exact quantum CD Hamiltonian $\hat H_{CD}(t)$~\cite{Deffner14}, in the form given by Eq.(\ref{HCD_scaleInvt}).

To investigate how well these results extend to systems that are driven in non-scale-invariant fashion, we will study a {\it tilted piston}: a particle of mass $m$ confined in a one-dimensional box with infinite walls and a slanted base. 
In terms of the length of the box $L$ and slope of its base $s$, the classical Hamiltonian is given by
\be
H_0 \left( q,p;s,L \right) = \frac{p^2}{2 m} + s q + \Theta(q;0,L),
\label{Hamiltonian}
\ee
where the function
\ba
\Theta(q;q_L,q_R) = 
\left\{
\begin{array}{lr}
0, & q_L < q < q_R\\
\infty & \text{otherwise}\\
\end{array}
\right.
\ea
describes hard walls at $q=0$ and $q=L$.

We subject the system to two different driving protocols. In case (a), the slope $s$ is changed while the length $L$ is held fixed, whereas in case (b), $s$ is 
held fixed and the box length is changed by moving the wall at $q=L$. Without loss of generality, we will assume that the slope is positive, $s>0$, and that the wall at 
$q=0$ remains fixed. Although the protocols (a) and (b) are both non-scale-invariant, when $s$ and $L$ are varied simultaneously while holding
$sL^3$ fixed, the system undergoes scale-invariant driving, as verified directly from Eqs.(\ref{eq:scaleInvt}) and (\ref{Hamiltonian}).

\section{Classical counterdiabatic terms} 
\label{sec:classical}

\begin{figure} 
\begin{center} 
\includegraphics[width=0.48\textwidth]{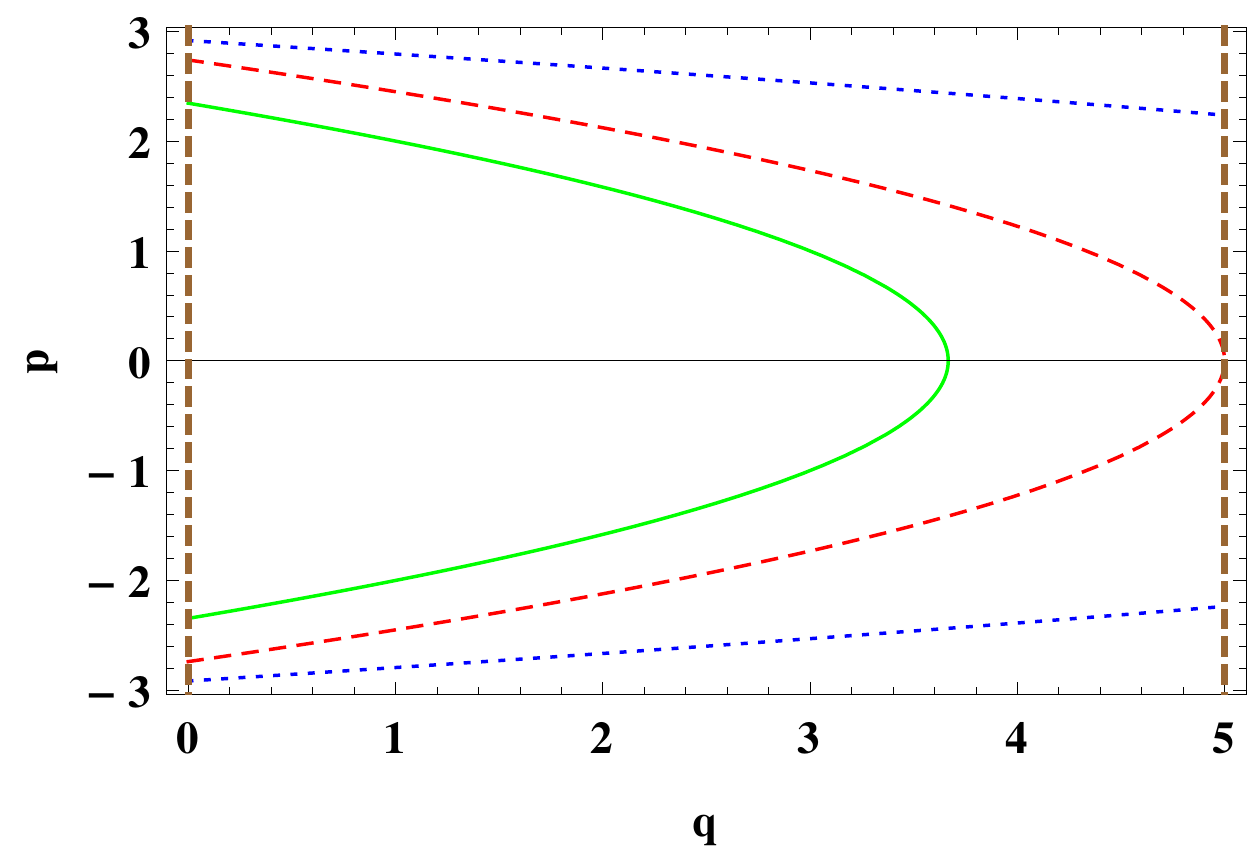} 
\caption{\label{Fig:ES} Three energy shells of $H_0$ (Eq.(\ref{Hamiltonian})) are shown for mass $m=1/2$, length $L=5$ and slope $s=1.5$. 
The green solid, red dashed and the blue dotted curves correspond to $E = 5.5 (< sL)$, $E =7.5 (= sL)$ and $E = 8.5 (> sL)$ respectively. The brown 
dashed lines at $q=0$ and $q=5$ denote the hard walls.} 
\end{center} 
\end{figure}

Fig.\ref{Fig:ES} illustrates the classical energy shells of the Hamiltonian $H_0(q,p;s,L)$ defined by Eq.(\ref{Hamiltonian}). There is a critial value of energy, 
$E_c = sL$, below which a classical particle interacts only with the wall at $q=0$, and above which the particle interacts with both walls. When $E \leq E_c$, the energy 
shell is a single curve that is symmetric about the $q$-axis and has a discontinuity at $q=0$, whereas when $E>E_c$, the energy shell is a pair of curves symmetric about 
the $q$-axis with discontinuities at $q=0$ and $q=L$. The expression for the classical CD generator $\boldsymbol{\xi}(z,\boldsymbol{\lambda}(t))$ depends on $E_c$. 

Let $\Omega_<$ denote the volume enclosed by an energy shell of energy $E<E_c$ and $\Omega_>$ denote the volume enclosed when $E>E_c$. 
From Eq.(\ref{Hamiltonian}) using $p=\sqrt{2 m (E-sq)}$ we obtain
\ba
\Omega_<(E,s,L) &=& 2 \int_0^{E/s} p \, dq = \frac{4 \sqrt{2m} E^{\frac{3}{2}}}{3 s},\label{O<} \\
\Omega_>(E,s,L) &=& 2 \int_0^{L} p \, dq = \frac{4 \sqrt{2 m}}{3 s} \left[  E^{\frac{3}{2}}-(E-sL)^{\frac{3}{2}} \right]. \non \\
\label{O>}
\ea

We now solve explicitly for the classical generator $\xi(z;\boldsymbol{\lambda})$ specified by Eqs.(\ref{xiexp}) and (\ref{xiavg}).
We analyze separately the two protocols (a) and (b) mentioned above.

\subsection{Case (a): time-dependent slope at fixed length}

The time-dependent parameter $\boldsymbol{\lambda}$ is the slope $s$ in this case.
Let us first solve for $\xi$ when $E \le E_c$. Using Eqs.(\ref{avgH}) and (\ref{O<})
\be
\bra \nabla H_0 \ket_{E,\lambda} = -\frac{\partial_s \Omega_<(E,s,L)}{\partial_E \Omega_<(E,s,L)} =\frac{2E}{3s},
\ee
and from Eq.(\ref{Hamiltonian})
\be
\nabla H_0 = \partial_s H_0 = q.
\label{dHq}
\ee
Hence, from Eq.(\ref{xiexp}), upon using $dt = m\,dq / \sqrt{2 m (E-sq)}$, we arrive at
\be
\xi(z_b;s)-\xi(z_b;s) = \int_a^b dt \nabla \tilde{H}_0(z(t);s) =-\frac{qp}{3s}\biggr\rvert_a^b.
 \label{<sl}
\ee
Together with the condition $\bra \xi \ket _{E,s}=0$, this result gives us, for $E<sL$,
\be
\xi_C^<(q,p;s) = -\frac{qp}{3s},
\label{xisll}
\ee
where the subscript $C$ stands for classical. This simple form for Eq.(\ref{xisll}) reflects the fact that when $E<sL$, varying the slope of the base at fixed box length 
is tantamount to scale invariant driving. 

We now construct the generator $\xi_C^>(q,p;s)$, for $E\ge E_c$. For the upper and the lower branches of the energy shell, let $\xi(0,p(0);s) = \xi_0$, and $\xi(L,p(L);s)
= \xi_1$ denote two constants to be determined later. Eqs.(\ref{avgH}) and (\ref{O>}) yield 
\be
\bra \partial_s H_0 \ket_{E,s} =\frac{E+ s L - \sqrt{E(E-sL)}}{3 s}.
\label{avgHq}
\ee
Using Eqs.(\ref{xiexp}), (\ref{dHq}) and (\ref{avgHq}), we obtain the following two equations for the upper and the lower branches
respectively:
\ba
&& \xi(z;s)-\xi_0 =  -\frac{p}{3s^2} \left[E- sL + \sqrt{E(E-sL)}\right] -\frac{pq}{3s} \non \\&&+  \frac{\sqrt{2mE}}{3s^2} \left[E- sL + \sqrt{E(E-sL)}\right],
\label{xi_s_u}
\ea
\ba
&& \xi(z;s)-\xi_1 = -\frac{p}{3s^2} \left[E- sL + \sqrt{E(E-sL)}\right] -\frac{pq}{3s}  \non \\ 
&& -\frac{\sqrt{2m(E-sL)}}{3s^2} \left[E + \sqrt{E(E-sL)}\right].
\label{xi_s_d}
\ea
The constants $\xi_0$ and $\xi_1$ are now determined by demanding continuity of $\xi$ at $q=0$ and $q=L$, along with the condition that $\bra \xi \ket_{E,s} =0$. A series
of simple calculations yield $\xi_0=0=\xi_1$. Eqs.(\ref{xi_s_u}) and (\ref{xi_s_d}) can be rearranged to obtain the general expression for the classical generator 
when $E>sL$:
\begin{widetext}
\be
\xi_C^>(q,p;s) = -\frac{p}{3s^2} \left[E- sL + \sqrt{E(E-sL)}\right] -\frac{pq}{3s} + \textrm{sign}(p) \cdot \frac{\sqrt{2m}}{3s^2} \left[E\sqrt{E-sL}+\sqrt{E}
(E-sL)\right],
\label{xislg}
\ee
\end{widetext}
where $\textrm{sign}(p)= +1$ for the upper branch and -1 for the lower branch.
As a consistency check, we note that at the critical energy $E=E_c=sL$, Eq.(\ref{xislg}) reduces to Eq.(\ref{xisll}).

\subsection{Case (b): time-dependent length at fixed slope}
In this case, the length $L$ plays the role of the parameter $\boldsymbol{\lambda}$. A particle with energy $E \le E_c$ is not influenced
by the motion of the wall at $q=L$. Hence we expect
\be
\xi_C^<(q,p;L)=0.
\label{xilnl}
\ee
This is mathematically verified by noting that $\partial_L H_0 = 0$ for $q<L$, and $\bra \partial_L H_0 \ket_{E,L} = 0$ (from Eqs.(\ref{avgH}) and (\ref{O<})), therefore 
Eq.(\ref{xiexp}) gives us $\xi_C^<=0$. 

When $E > E_c$, Eqs.(\ref{avgH}) and (\ref{O>}) yield $\bra \partial_L H_0 \ket_{E,L} = -\left[E-sL+ \sqrt{E(E-sL)}\right]/L$. Hence at all points except at $q=L$, 
\be
\partial_L \tilde{H}_0(z(t);L) = \frac{E-sL+ \sqrt{E(E-sL)}}{L}.
\label{deltildeHL}
\ee
Analogous to case (a), to-be-determined constants $\xi(0,p(0);L) = \xi_0'$ and $\xi(L,p(L);L) = \xi_1'$ are introduced for the upper and lower branches respectively. 
Using Eqs.(\ref{xiexp}) and (\ref{deltildeHL}) we obtain, for the upper and the lower branches 
respectively:
\ba
\xi(z;L)-\xi_0' = \frac{ E- sL + \sqrt{E(E-sL)}}{sL} \left[\sqrt{2mE} -p\right],\non \\
\label{xi_l_u}
\ea
\ba
\xi(z;L)-\xi_1' &=&\frac{ E- sL + \sqrt{E(E-sL)}}{sL} \times \non \\ && \left[-\sqrt{2m(E-SL)} -p \right].\non \\
\label{xi_l_d}
\ea
Setting $\bra \xi \ket_{E,L} =0$ and demanding continuity of $\xi$ at $q=0$,  we get $\xi_0'=0$ and 
\ba
\xi_1'&=&\frac{ E- sL + \sqrt{E(E-sL)}}{sL} \times  \non \\ &&\left[\sqrt{2m(E-SL)} -\sqrt{2mE}\right]. \non \\
\ea
Eqs.(\ref{xi_l_u}) and (\ref{xi_l_d}) can be combined to give the classical generator for $E > E_c$:
\begin{widetext}
\be
\xi_C^>(q,p;L) = -\frac{p}{sL} \left[E- sL + \sqrt{E(E-sL)}\right] + \textrm{sign}(p) \cdot \frac{\sqrt{2m}}{sL} \left[E\sqrt{E-sL}+\sqrt{E}(E-sL)\right],
\label{xilng}
\ee
\end{widetext}
which is consistent with Eq.\ (\ref{xilnl}) at $E=E_c$. 

Eqs.(\ref{xisll}) and (\ref{xilnl}) provide explicit expression for $\xi$ at energies $E<E_c$, and Eqs.(\ref{xislg}) and (\ref{xilng}) give $\xi$ for $E\ge E_c$.
As mentioned earlier, below the critical energy the system is effectively driven in a scale-invariant manner.
We will focus our attention on energies above the critical energy, where the driving is non-scale-invariant.

Comparing Eqs.(\ref{xisll}) and (\ref{xislg}) with Eqs.(\ref{xilnl}) and (\ref{xilng}) respectively, we note that the classical generators for cases (a) and (b) are 
related to each other by the following relation:
\be
\xi_C(q,p;s) +\frac{p q}{3s} = \xi_C(q,p;L) \cdot \frac{L}{3s}.
\label{relnc}
\ee

\section{Semiclassical counterdiabatic terms}
\label{sec:sc}

Having obtained exact classical expressions for the generator $\xi(z;\boldsymbol{\lambda})$, we now wish to utilize these results to construct its quantum counterpart 
$\hat\xi(\boldsymbol{\lambda})$, in terms of position and momentum operators $\hat{q}$ and $\hat{p}$. 
In later sections we will study, numerically, the extent to which the operator constructed in this manner produces transitionless quantum driving for the quantum tilted
piston.

We seek  a semiclassical approximation for the quantum generator, 
denoted by $\hat{\xi}_{SC}$.
In cases (a) and (b) described above, $\hat{H}_{CD}(t)$ is given by $\dot{s} \cdot \hat{\xi}_{SC}(q,p;s)$ and $\dot{L} \cdot \hat{\xi}_{SC}(q,p;L)$ respectively, where 
$\hat{\xi}_{SC}$ is Hermitian. As the operators $\hat{q}$ and $\hat{p}$ do not commute, merely putting `hats' on the observables in 
Eqs.(\ref{xisll}), (\ref{xislg}) and (\ref{xilng}) will not ensure Hermiticity. Rather, the terms in $\xi_C$ must be symmetrized. Complete 
symmetrization as prescribed in Ref.\cite{Julg75} becomes unfeasible as $\xi_C$ contains terms with non-integer powers of 
$q$ and $p$. We therefore implement the following procedure to symmetrize the expressions.

Any term in $\xi_C$ of the form $f(p) \cdot g(E)$, where $f$ and $g$ are 
arbitrary functions, is symmetrized as
\be
\frac{f(\hat{p}) \cdot g(\hat{H_0}) + g(\hat{H_0}) \cdot f(\hat{p})}{2},
\ee
where $\hat{H_0}$ is the quantized version of Eq.(\ref{Hamiltonian}). The semiclassical operators for $E \le sL$ are given by
\ba
\hat{\xi}_{SC}^<(\hat{q},\hat{p};s)&=&-\frac{1}{6s}(\hat{q}\hat{p}+\hat{p}\hat{q}), \non \\
\hat{\xi}_{SC}^<(\hat{q},\hat{p};L)&=&0,
\label{xil}
\ea
 where Eqs.(\ref{xisll}) and (\ref{xilnl}) have been used. Let $\hat{\eta}$ be the semiclassical operator for sign($p$) (to be determined later). 
For $E > sL$, from Eq.(\ref{xislg}) and (\ref{xilng}), we obtain
\begin{subequations}
\label{eq:scHCD}
\ba
\hat{\xi}_{SC}^>(\hat{q},\hat{p};s) &=& -\frac{1}{3s^2} \hat{\xi}_{1} -\frac{1}{3s} \hat{\xi}_{2}+\frac{1}{3s^2} \hat{\xi}_{3}, \non \\
\hat{\xi}_{SC}^>(\hat{q},\hat{p};L) &=&  -\frac{1}{sL} \hat{\xi}_{1} +\frac{1}{sL} \hat{\xi}_{3},
\label{xig}
\ea
where
\ba
\label{xi1}
\hat{\xi}_{1} &=&  \frac{\hat{p} \cdot f(\hat{H_0}) + f(\hat{H_0}) \cdot \hat{p}}{2} \\
\label{xi2}
\hat{\xi}_{2} &=& \frac{\hat{q}\hat{p}+\hat{p}\hat{q}}{2} \\
\label{xi3}
\hat{\xi}_{3} &=& \frac{\hat{\eta} \cdot g(\hat{H_0}) + g(\hat{H_0}) \cdot \hat{\eta}}{2} \\
\label{fexp}
f(\hat{H_0}) &=& \hat{H_0}- sL + \sqrt{\hat{H_0}(\hat{H_0}-sL)} \\
g(\hat{H_0}) &=& \sqrt{2m}\left[\hat{H_0}\sqrt{\hat{H_0}-sL}+\sqrt{\hat{H_0}}(\hat{H_0}-sL)\right] \nonumber\\
\label{gexp}
\ea
\end{subequations}

The generators $\hat{\xi}_{SC}(\hat{q},\hat{p};s)$ and $\hat{\xi}_{SC}(\hat{q},\hat{p};L)$ defined by 
Eqs.(\ref{xil})-(\ref{xi3}) satisfy
\be
\hat{\xi}_{SC}(\hat{q},\hat{p};s) + \frac{1}{3s} \cdot \frac{\hat{q}\hat{p}+\hat{p}\hat{q}}{2}= \hat{\xi}_{SC}(\hat{q},\hat{p};L) \cdot \frac{L}{3s} ,
\label{relnsc}
\ee
which is the semiclassical counterpart of Eq.(\ref{relnc}). 

\section{Solving the time-dependent Schr\"odinger Equation}
\label{sec:solving}

In the previous section, we obtained semiclassical expressions for the generators $\hat{\xi}_{SC}(\hat{q},\hat{p};\boldsymbol{\lambda})$, where $\boldsymbol{\lambda}=s$ 
for case (a) and $\boldsymbol{\lambda}=L$ for case (b). We now aim to simulate the evolution of the system under the time-dependent Schr\"odinger equation (TDSE), to 
establish how well these generators produce transitionless quantum driving. 

Let the wavefunction
\be
u_n(q,\boldsymbol{\lambda}) = \langle q \vert n(\boldsymbol{\lambda}) \rangle
\ee
denote the $n^{th}$ eigenstate of the unperturbed Hamiltonian $\hat{H}_0(\boldsymbol{\lambda})$, in the position representation.
For a given protocol $\boldsymbol{\lambda}(t)$, we will evolve a wavefunction $\psi(q,t)$ under the TDSE, 
${\mathcal H} \psi = i\hbar \, \partial\psi/\partial t$, using the Hamiltonian 
\be
\hat{\mathcal{H}}(t) = \hat{H}_0(\boldsymbol{\lambda}(t)) +\dot{\boldsymbol{\lambda}} \cdot \hat{\xi}_{SC}(\hat{q},\hat{p}; \boldsymbol{\lambda}(t)),
\label{SCSEqn}
\ee
with initial condition $\psi(q,0) = u_n(q,\boldsymbol{\lambda}(0))$.
We will compare the evolving wavefunction $\psi(q,t)$ with the instantaneous $n^{th}$ energy eigenstate by evaluating the \textit{fidelity}:
\be
\mathcal{F}(t)=|\bra n(\boldsymbol{\lambda}(t)) \vert \psi(t)\ket| = \int dq\, u_n^*(q,\boldsymbol{\lambda}(t)) \psi(q,t)
\ee
The fidelity provides a direct measure, between $0$ and $1$, of the degree to which the term $\dot{\boldsymbol{\lambda}} \cdot \hat{\xi}_{SC}$ appearing in
Eq.\ (\ref{SCSEqn}) suppresses transitions out of the $n^{th}$ energy eigenstate.

In this section we describe our approach to solving the TDSE numerically, and we develop the tools required to implement this procedure.
We expand the time-dependent wave function as
\be
\psi(q,t)=\sum\limits_{n}  a_n(t) u_n(q,\boldsymbol{\lambda}(t)) \exp \left[-\frac{i}{\hbar} \int_0^t E_n(t')dt' \right],
\label{wf}
\ee
where $E_n(t)$ is the $n^{th}$ eigenvalue of $\hat{H}_0(\boldsymbol{\lambda}(t))$, and the expansion coefficients satisfy $\sum_n |a_n(t)|^2=1$.
Upon substituting Eq.(\ref{wf}) in the TDSE, using the Hamiltonian $\hat{\mathcal{H}}(t)$ given by Eq.(\ref{SCSEqn}), we obtain
\be
\dot{a}_m=\sum\limits_{n} N_{mn}a_n,
\label{deqn}
\ee
where
\be
 N_{mn} = \dot{\boldsymbol{\lambda}} \exp \left[-\frac{i}{\hbar} \int_0^t (E_n(t')-E_m(t'))dt' \right] M_{mn},
 \label{nmeqn}
\ee
 and 
\ba
M_{mn} &=& - \bra m| \boldsymbol{\nabla} n \ket  + \frac{1}{i\hbar} \bra m|\hat{\xi}_{SC} |n \ket \non \\
  &=& M^0_{mn} + M^{CD}_{mn}.
  \label{meqn}
\ea
The term $M^0_{mn}\equiv - \bra m| \boldsymbol{\nabla} n \ket \non$ arises from the term $\hat{H}_0$ in Eq.(\ref{SCSEqn}), while
 $M^{CD}_{mn}\equiv(i\hbar)^{-1}\bra m| \hat{\xi}_{SC} |n \ket$ is the contribution from the semiclassical CD generator, $\dot{\boldsymbol{\lambda}} \cdot \hat{\xi}_{SC}$.  Solving the Schr\"odinger equation is equivalent
 to solving the first order matrix differential equation Eq.(\ref{deqn}) for the expansion coefficients $a_n(t)$.
 
 In order to obtain explicit expressions for the matrices $M^0$ and $M^{CD}$ appearing in Eq.(\ref{meqn}), it is convenient to make use of two different time-dependent basis sets in Hilbert space.
 The first is the {\it energy basis}, $\{ \vert n(\boldsymbol{\lambda}) \ket \}$, consisting of the eigenstates of $\hat H_0(\boldsymbol{\lambda})$.
 The second is the {\it sine basis}, $\{ \vert \alpha(L) \ket \}$, by which we mean the orthogonal sinusoidal functions of length $L$:
\be
\label{eq:sineBasis}
\bra q | \alpha(L) \ket = \sqrt{\frac{2}{L}} \sin \left( \frac{\alpha \pi q}{L} \right) \quad,\quad \alpha\ge 1
\ee
where $L=L(t)$.
We will use Latin and Greek letters, respectively, to denote energy and sine basis states.
Given a Hermitian operator $\hat{O}$, its representation in the energy and sine bases will be denoted by the matrices
\begin{subequations}
\ba
\bar{O}_{mn}&=&\bra m| \hat{O} | n \ket  \\
\tilde{O}_{\alpha \beta}&=&\bra \alpha | \hat{O} | \beta \ket
\ea
\end{subequations}

The operators $f(\hat{H_0})$ and $g(\hat{H_0})$, defined by Eqs.(\ref{fexp}) and (\ref{gexp}), are conveniently represented in the energy basis, in which they become diagonal matrices with entries
 \ba
 \label{eq:fmm}
 \bar f_{mm} &=& E_m- sL + \sqrt{E_m(E_m-sL)} \\
 \label{eq:gmm}
 \bar g_{mm} &=& \sqrt{2m}\left[E_m\sqrt{E_m-sL}+\sqrt{E_m}(E_m-sL)\right] \nonumber\\
 \ea
 
The operators $\hat{p}$, $\hat{\xi}_2$ and $\hat{\eta}$ are more conveniently represented in the sine basis.
Using Eq.(\ref{eq:sineBasis}), we obtain
\ba
\label{eq:ptilde}
\tilde p_{\alpha\beta} = 
\left\{
\begin{array}{ll}
0 & \quad \alpha -\beta = even\\
\frac{4 i\hbar \alpha \beta}{L(\beta^2 - \alpha^2)} & \quad \alpha -\beta = odd
\end{array},
\right.
\ea
and
\ba
(\tilde\xi_2)_{\alpha\beta} =
\left\{
\begin{array}{ll}
0 & \quad \alpha = \beta\\
-\frac{2 i \hbar \alpha \beta}{\beta^2 - \alpha ^2} & \quad \alpha -\beta = even \, , \, \alpha\ne\beta \\
\frac{2 i \hbar \alpha \beta}{\beta^2 - \alpha ^2} & \quad \alpha -\beta = odd
\end{array}
\right.
\label{qpmatrix}
\ea
A representation of $\hat{\eta}$ in the sine basis is obtained by semiclassical means in the Appendix, yielding the result:
\ba
\label{eq:etatilde}
\tilde{\eta}_{\alpha \beta} =
\left\{
\begin{array}{ll}
0 & \quad \alpha -\beta = even\\
\frac{2 i}{(\beta-\alpha) \pi} & \quad \alpha -\beta = odd\\
\end{array}
\right.
\label{eta}.
\ea

In order to use Eqs.~\ref{eq:fmm} - \ref{eq:etatilde} to construct the matrix elements $M^{CD}_{mn}$, we required the similarity transformation
\ba
\label{eq:similarity}
\bar{O}_{mn} = \sum_{\alpha \beta} Z^{\dagger}_{m \alpha} \tilde{O}_{\alpha \beta} Z_{\beta n},
\ea
where $Z_{\beta n}=\bra \beta | n \ket$.
$Z$ is the matrix that diagonalizes $\tilde{H}_0$ -- the sine basis representation of $\hat{H}_0$ -- which can be evaluated explicitly: 
\ba
(\tilde{H}_0)_{\alpha \beta} =
\left\{
\begin{array}{ll}
0 & \alpha -\beta = even \, , \, \alpha\ne\beta\\
-\frac{8 \alpha \beta s L}{(\alpha ^2 - \beta ^2)^2 \pi^2} & \alpha -\beta = odd\\
\frac{(\alpha \pi \hbar)^2}{2 m L^2} + \frac{s L}{2} & \alpha = \beta
\end{array}.
\right.
\ea
We obtained $Z$ from $\tilde{H}_0$ numerically, and we used the result to transform $\tilde p$, $\tilde\xi_2$ and $\tilde\eta$ (Eqs.(\ref{eq:ptilde} - \ref{eq:etatilde})) into $\bar p$, $\bar\xi_2$ and $\bar\eta$ via Eq.(\ref{eq:similarity}).
We then combined these expressions with $\bar f$ and $\bar g$ (Eqs.(\ref{eq:fmm}), (\ref{eq:gmm})) to construct $\bar\xi_1$, $\bar\xi_2$ and $\bar\xi_3$ (see Sec.\ \ref{sec:sc}).
Finally, from these we obtained $\bar\xi_{SC}$ and therefore $M^{CD}_{mn}$ (Eq.(\ref{meqn})).

 
In addition to $M^{CD}_{mn}$, Eq.(\ref{meqn}) contains the term
\be
M^0_{mn} = - \bra m | \nabla n \ket = -\frac{\bra m | \nabla \hat{H_0}|n \ket}{E_n-E_m}
\ee
For case (a), $\nabla \hat{H_0}=\partial_s \hat{H_0}=\hat{q}$.
The elements of $\hat{q}$ in the sine basis are
\ba
\tilde{Q}_{\alpha \beta} =
\left\{
\begin{array}{ll}
0 & \alpha -\beta = even \, , \, \alpha\ne\beta \\
-\frac{8 \alpha \beta L}{(\alpha ^2 - \beta ^2)^2 \pi^2} & \alpha -\beta = odd\\
\frac{L}{2} & \alpha = \beta
\end{array}.
\right.
\ea
After obtaining $\bar{Q} = Z^T \tilde{Q} Z$, we have
\ba
- \bra m | \partial_s n \ket =
\left\{
\begin{array}{lr}
- \frac{\bar{Q}_{mn}}{E_n-E_m} & m \neq n \\
0 & m=n
\end{array}.
\right.
\label{mgradnsl}
\ea

For case (b) we have $\nabla \hat{H_0}=\partial_L \hat{H_0}$, whose classical counterpart $\partial_L H_0$ is singular at $q=L(t)$. We will determine $\bra m | \partial_L n \ket$ by relating it to $\bra m | \partial_s n \ket$ using scale invariance.

The potential $V(q;s,L) = sq + \Theta(q;0,L)$ that appears in our Hamiltonian, Eq.(\ref{Hamiltonian}), depends parametrically on both the slope $s$ and the length $L$.
If these two parameters are constrained to satisfy
\be
\label{sl}
s(L)L^3 = constant
\ee
(treating the slope $s$ as a function of the length $L$) then the potential function satisfies
\be
\label{scale} 
V(q;s(L),L) = \frac{1}{L^2} V \left( \frac{q}{L};s(1),1 \right)
\ee
which is the condition for scale invariance.
In this situation the $n^{th}$ energy eigenfunction satisfies \cite{Deffner14}
\be
 \label{wavefunc}
 u_m(q;s(L),L) = \frac{1}{\sqrt{L}} u_m \left( \frac{q}{L};s(1),1 \right). 
\ee
Differentiating both sides of Eq.(\ref{wavefunc}) with respect to $L$ and equating the results, we get
\be
 \frac{\partial u_m}{\partial L} = \frac{ 3 s}{ L} \frac{\partial u_m}{\partial s} -\frac{u_m}{2L}-\frac{q}{L}\frac{\partial u_m}{\partial q}.
 \label{eqnbl}
\ee
Since the entire parameter space can be filled by a set of non-intersecting curves defined by Eq.(\ref{sl}), Eq.(\ref{eqnbl}) is valid for any slope $s$ and any positive length $L$.
 
Now consider the expressions
\ba
\bra n | \partial_L m \ket &=& \int_0^L dq\  u_n^* \frac{\partial u_m(q)}{\partial L} \non \\
&=&  \frac{ 3 s}{ L} \bra n | \partial_s m \ket - \frac{1}{2L} \delta_{mn}  - \non \\ && \frac{1}{L}
\int_0^L dq \  u_n^* \ q \frac{\partial u_m(q)}{\partial q} 
\label{mgradn}
\ea
 and
\ba
\frac{1}{i \hbar} \bra n|\frac{\hat{q} \hat{p} + \hat{p} \hat{q}}{2}|m \ket &=& 
-\frac{1}{2}  \int_0^L dq \ u_n^* \ q \frac{\partial u_m(q)}{\partial q}\non \\
&& - \frac{1}{2}\int_0^L dq \  u_n^* \ \frac{\partial (q \ u_m(q))}{\partial q} \non \\
&=& - \int_0^L dq \  u_n^* \ q \frac{\partial u_m(q)}{\partial q} -\frac{\delta _{mn}}{2}\non \\
\label{xp}
\ea
Substituting Eq.(\ref{xp}) into Eq.(\ref{mgradn}), we arrive at
\be
\bra n | \partial_L m \ket= \frac{ 3 s}{ L} \bra n | \partial_s m \ket + 
\frac{1}{L} \frac{1}{i \hbar} \bra n|\frac{\hat{q} \hat{p} + \hat{p} \hat{q}}{2}|m \ket,
\label{relation}
\ee
which can alternatively be obtained from Eq.(\ref{relnsc}). We can therefore compute the matrix representation of $\bra n | \partial_L m \ket$ as we
have already determined both the terms on the right side of Eq.(\ref{relation}): Eq.(\ref{mgradnsl}) gives the first term while the second is obtained after performing a similarity transformation on the matrix given in Eq.(\ref{qpmatrix}). 

\section{Numerical Results}

\begin{figure*} 
\begin{center} 
\includegraphics[width=0.75\textwidth]{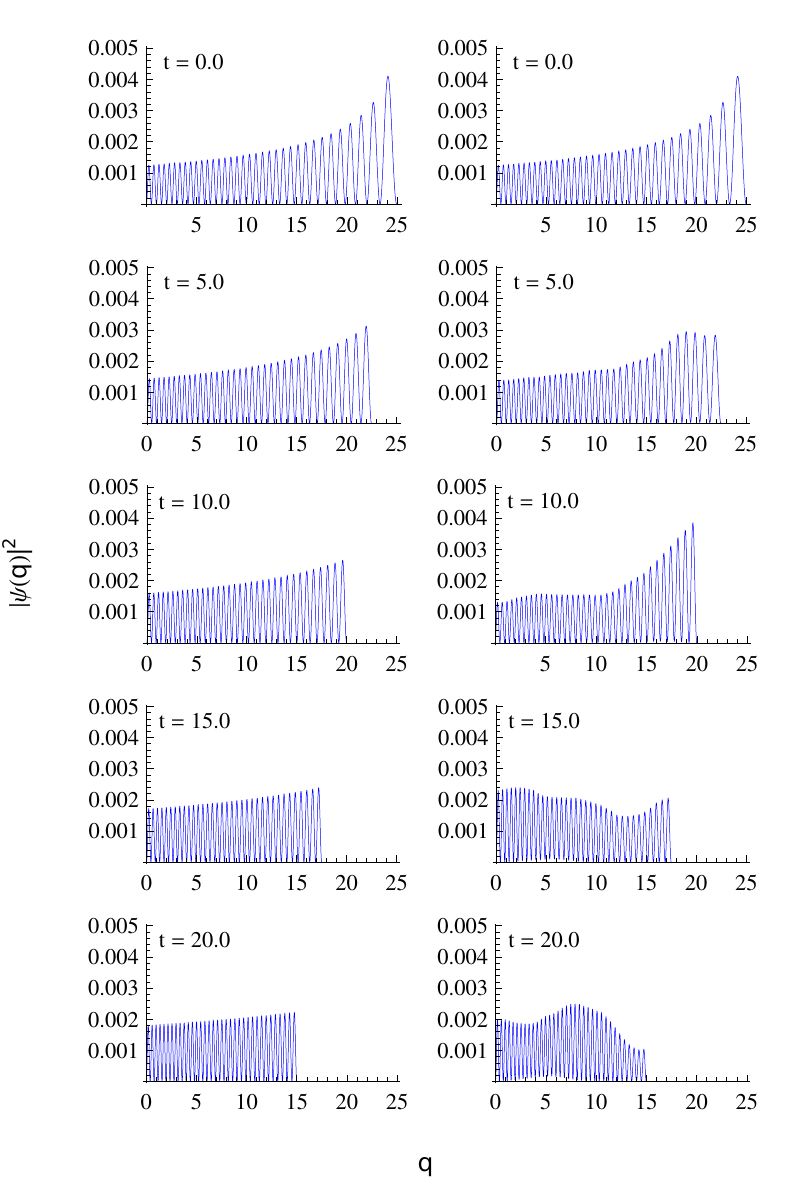} 
\caption{\label{Fig:pdl} Evolution of the probability density $|\psi(q,t)|^2$ for a particle of mass $m=1.0$ in a box whose slope is
fixed at $s=3.0$ and whose length is decreased from $L=25.0$ to $15.0$ at a rate $\dot{L}=-0.5$. Snapshots of the wavefunction are taken at times t=0, 
5.0, 10.0, 15.0 and 20.0. The plots on the left depict evolution under the full Hamiltonian $\hat{\mathcal{H}}(t) = \hat{H}_0(t) + \dot{\boldsymbol{\lambda}} \cdot 
\hat{\xi}_{SC}(\boldsymbol{\lambda}(t))$, while those on the right depict evolution under  $\hat{H}_0(t)$.}
\end{center} 
\end{figure*} 

\begin{figure*} 
\begin{center} 
\includegraphics[width=0.75\textwidth]{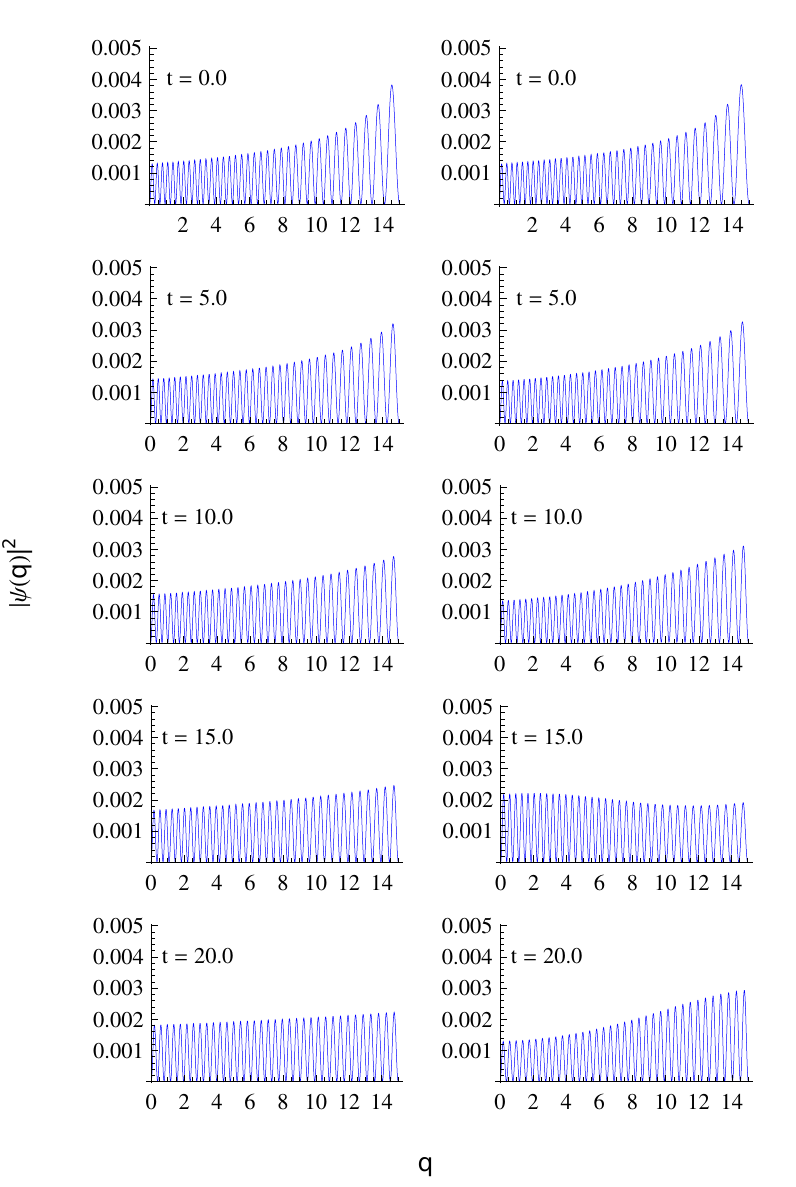} 
\caption{\label{Fig:pds}Same as Fig.\ref{Fig:pdl} except that the length of the box is fixed at $L=15.0$, while the slope is decreased from 
$s=13.0$ to $3.0$ at a rate $\dot{s}=-0.5$.} 
\end{center} 
\end{figure*} 

\begin{figure*} 
\begin{center} 
\includegraphics[width=0.8\textwidth]{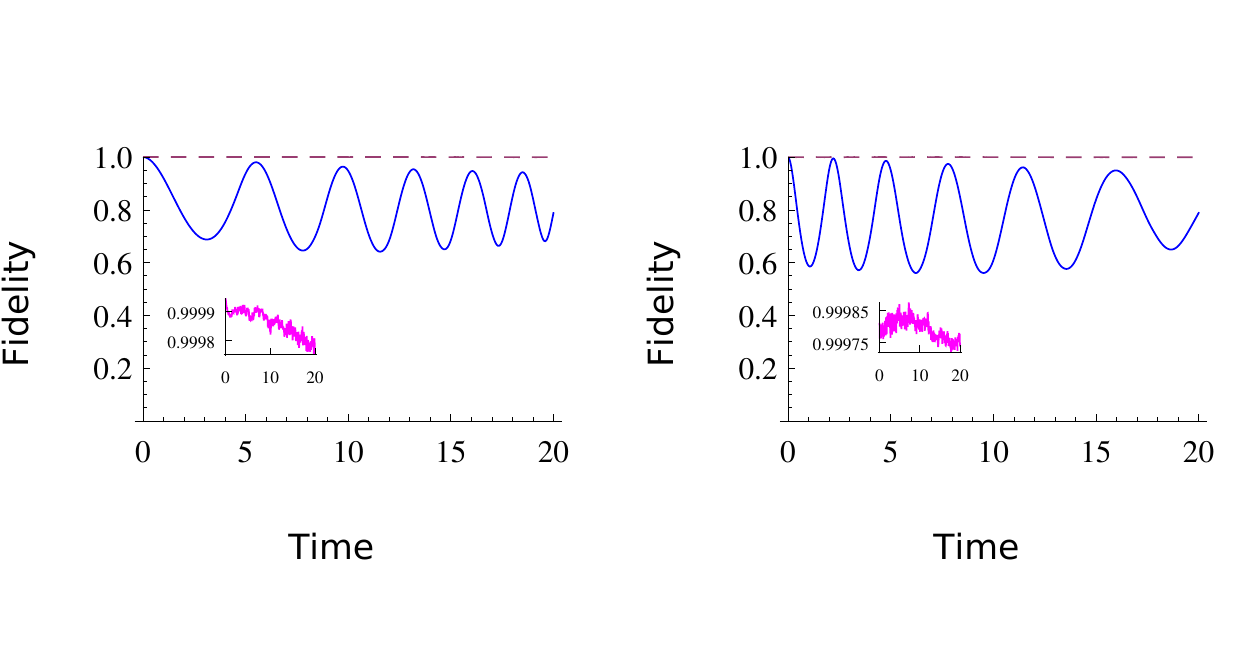} 
\caption{ \label{Fig:Fidlen} Evolution of the fidelity ${\mathcal F}(t)$. The plot on the left is for the case shown in Fig.\ref{Fig:pdl}, whereas the plot on the 
right is for the same system but subjected to the reverse process: the box length increases from $L=15.0$ to $25.0$ at $\dot{L}=0.5$.
The dashed magenta curve depicts the fidelity for evolution under $\hat{\mathcal{H}} = \hat{H}_0 + \dot{\boldsymbol{\lambda}}
\cdot \hat{\xi}_{SC}$, while the blue curve is the fidelity upon evolution under $\hat{H}_0$. The inset is a magnified 
view of the dashed magenta curve.} 
\end{center} 
\end{figure*} 

\begin{figure*} 
\begin{center} 
\includegraphics[width=0.8\textwidth]{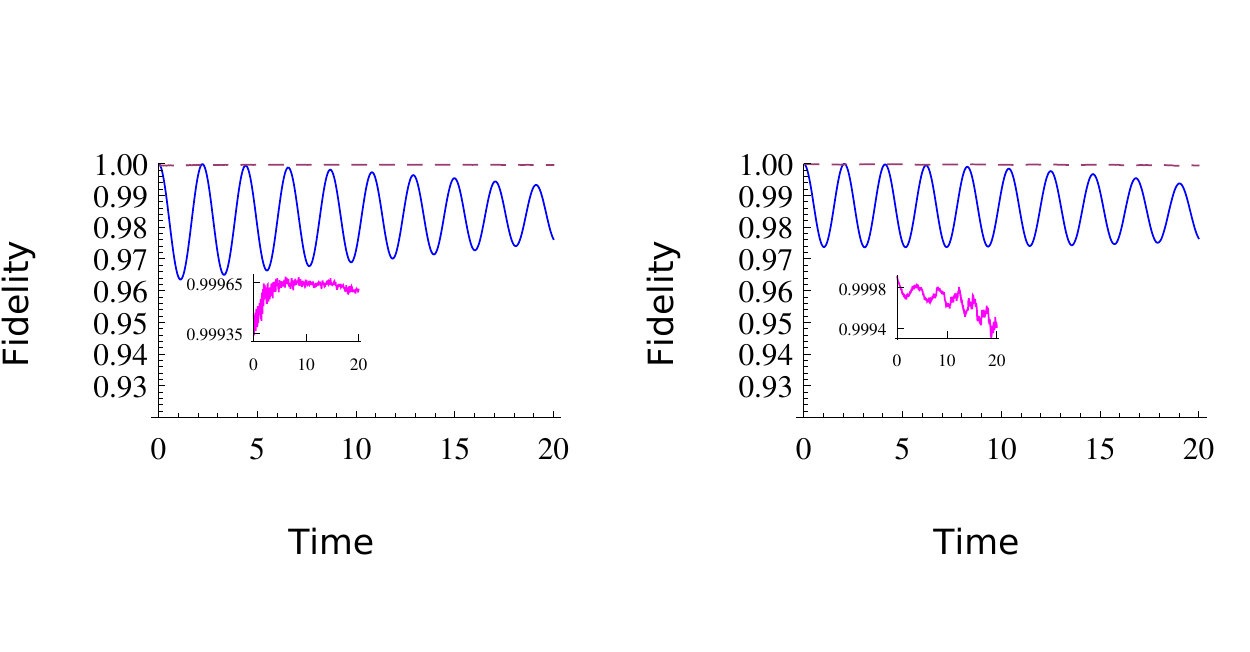} 
\caption{\label{Fig:FidSlp} Similar to Fig.\ref{Fig:Fidlen}. The left plot is for the case shown in Fig.\ref{Fig:pds}, whereas in the right plot
the same system is subjected to the reverse process: the slope $s$ increases from $s=3.0$ to $13.0$ at $\dot{s}=0.5$.} 
\end{center} 
\end{figure*}

Having determined the matrices $M^0$ and $M^{CD}$ (Eq.(\ref{meqn})), we solved the time-dependent Schr\"odinger equation by numerically integrating Eq.(\ref{deqn}) using the (fourth-order) Runge-Kutta-Gill  method \cite{Gill51}.
In each simulation the system was initialized in the $n^{th}$ energy eigenstate, $a_k(0) = \delta_{kn}$, then it was evolved in time as either the slope $s$ was varied at fixed length $L$ (case (a)), or else the length was varied at constant slope (case (b)).
The rate of change, $\dot s$ or $\dot L$, was set to a constant value sufficiently large to produce non-adiabatic evolution in the absence of the counterdiabatic term.

Simulations were performed both under the original Hamiltonian $\hat{H}_0(t)$, describing the time-dependent tilted piston (Eq.(\ref{Hamiltonian})), and under the composite Hamiltonian
\be
\hat{\mathcal{H}}(t) = \hat{H}_0(t) + \dot{\boldsymbol{\lambda}} \cdot \hat{\xi}_{SC}(\boldsymbol{\lambda}(t))
\ee
that includes the counterdiabatic term.
In both cases the fidelity $\mathcal{F}(t)=|\bra n(\boldsymbol{\lambda}(t)) | \psi(t) \ket|$ was computed.
In these simulations the particle mass was set to $m=1$ and Planck's reduced constant to $\hbar = 2$, and the system was initialized in the quantum number $n=35$.
The results, Figs.~\ref{Fig:pdl} - \ref{Fig:FidSlp}, are discussed in the following paragraphs.


Fig.~\ref{Fig:pdl} shows the evolving probability distribution $|\psi(q,t)|^2$ as the length of the tilted piston is reduced from $L=25.0$ to $15.0$, at a rate $\dot{L} = -0.5$ and fixed slope $s=3.0$.
The left column shows snapshots of $|\psi|^2$ at five instants in time, for evolution under the Hamiltonian $\hat{\mathcal{H}}(t)$.
The right column shows evolution under $\hat{H}_0(t)$.
In these simulations the initial energy is $E_n = 79.52$.
The plots on the left are visually 
indistinguishable from the probability distribution of the adiabatic energy eigenstate, $\vert u_n(q,\boldsymbol{\lambda}(t)) \vert^2$, with $n=35$.
By contrast, in the plots on the right the probability distribution develops noticeable shock waves, due to the rapid compression of the piston length.
Thus, with the addition of the counterdiabatic term the system faithfully follows a fixed eigenstate of $\hat{H}_0(t)$ (left plots), while in the absence of this term it is unable to keep pace with the rapidly changing Hamiltonian (right plots).

Fig.\ref{Fig:pds} presents evolution in a tilted piston of fixed length $L=15.0$, with a slope that decreases from $s=13.0$ to $3.0$ at a rate $\dot{s}=-0.5$.
As in Fig.~\ref{Fig:pdl}, the plots in the left and right columns depict evolution with and without the counterdiabatic term $\dot{\boldsymbol{\lambda}} \cdot \hat{\xi}_{SC}$.
Once again, the plots on the left are indistinguishable from the instantaneous energy eigenstate $\vert u_n(q,\boldsymbol{\lambda}(t)) \vert^2$, while those on the right reveal (mild) shock waves that are evidence of non-adiabatic evolution.
The counterdiabatic term again successfully guides the wavefunction along the desired adiabatic trajectory.

These claims are supported by analyses of the fidelity $\mathcal{F}(t)$. 
Fig.\ref{Fig:Fidlen} shows fidelity plots for a tilted piston undergoing compression (left plot) and expansion (right plot).
The former corresponds to the evolution shown in Fig.\ref{Fig:pdl}, while the latter depicts the reverse process, in which the length {\it increases} from $L=15.0$ to $25.0$ at $\dot{L}=0.5$.
Similarly, Fig.\ref{Fig:FidSlp} shows a fidelity plot for the evolution depicted in Fig.\ref{Fig:pds} (left plot), and for the reverse process in which the slope is varied from $s=3.0$ to $13.0$ at $\dot{s}=0.5$ (right plot).
In these figures, the solid blue curves depict the fidelity for evolution under $\hat{H}_0(t)$, while the dashed magenta curves correspond to evolution under 
 $\hat{\mathcal{H}}(t)$. In all four plots the blue curves deviate significantly, while the dashed magenta curves remain very close to unity, confirming that our semiclassically obtained counterdiabatic term has the desired effect of enforcing adiabatic evolution, with high accuracy.
 
 As a side comment we observe that, in Fig.\ref{Fig:Fidlen}, the oscillations in $\mathcal{F}(t)$ become more rapid in time when the tilted piston is compressed (left plot), and less rapid as it expands (right plot).
 These oscillations reflect the shock waves propagating between the two walls of the box, hence it makes sense that the period of oscillation diminishes or grows as the length $L$ decreases or increases.
  
 Because the counterdiabatic term $\dot{\boldsymbol{\lambda}} \cdot \hat{\xi}_{SC}$ was obtained semiclassically, we expect its efficacy to degrade as we approach the deep quantum regime.
 To test this hypothesis, we performed simulations at fixed slope $s=3.0$, with piston length decreasing from $L=25.0$ to $15.0$ at $\dot{L}=0.5$, and with particle mass $m=1$, as in Fig.\ref{Fig:pdl}.
 We carried out seven such simulations, with the value of $\hbar$ ranging from 1.0 to 7.0, choosing the initial state $n$ so that the particle starts with energy $E_n \approx 80$ in each simulation.
 Thus Planck's constant was varied while the classical parameters remained essentially fixed.
 As before, the system was subjected to evolution under both $\hat{H}_0(t)$ and $\hat{\mathcal{H}}(t)$, and the fidelity $\mathcal{F}(t)$ was computed.
Table \ref{table} lists $\mathcal{F}^{wcd}_{min}$, which is the minimum fidelity (over the duration of the process) when the system evolves under $\hat{\mathcal{H}}(t)$, and 
 $\mathcal{F}^{wocd}_{min}$, the minimum fidelity when the system evolves under $\hat{H}_0(t)$. 
 We see that as $\hbar$ increases and $n$ decreases -- that is, as we go deeper into the quantum regime -- $\mathcal{F}^{wcd}_{min}$ deviates further from unity.
 As expected, the semiclassical counterdiabatic term $\dot{\boldsymbol{\lambda}} \cdot \hat{\xi}_{SC}$ works best in the semiclassical limit of small $\hbar$ / large $n$.
 
Interestingly, Table \ref{table} reveals that $\mathcal{F}^{wocd}_{min}$ {\it increases} with $\hbar$: in the absence of the counterdiabatic term, the fidelity improves as we go deeper into the quantum regime.
We attribute this behavior to the fact that the spacing between adjacent energy levels increases with $\hbar$.
Let us picture a classical process in which initial conditions are sampled from a single energy shell, and trajectories evolve from these initial conditions under the Hamiltonian $H_0(q,p,t)$.
The final distribution of energies then provides a crude estimate of the final energy distribution in the corresponding quantum process, in which the system begins in an energy eigenstate.
For a fixed final distribution of energies, the distribution of final quantum numbers decreases with increasing $\hbar$, simply because of the decreasing density of energy levels.
As a result, excitations out of the initial energy level are inhibited. 

\begin{figure} 
\centering 
\includegraphics[width=0.4\textwidth]{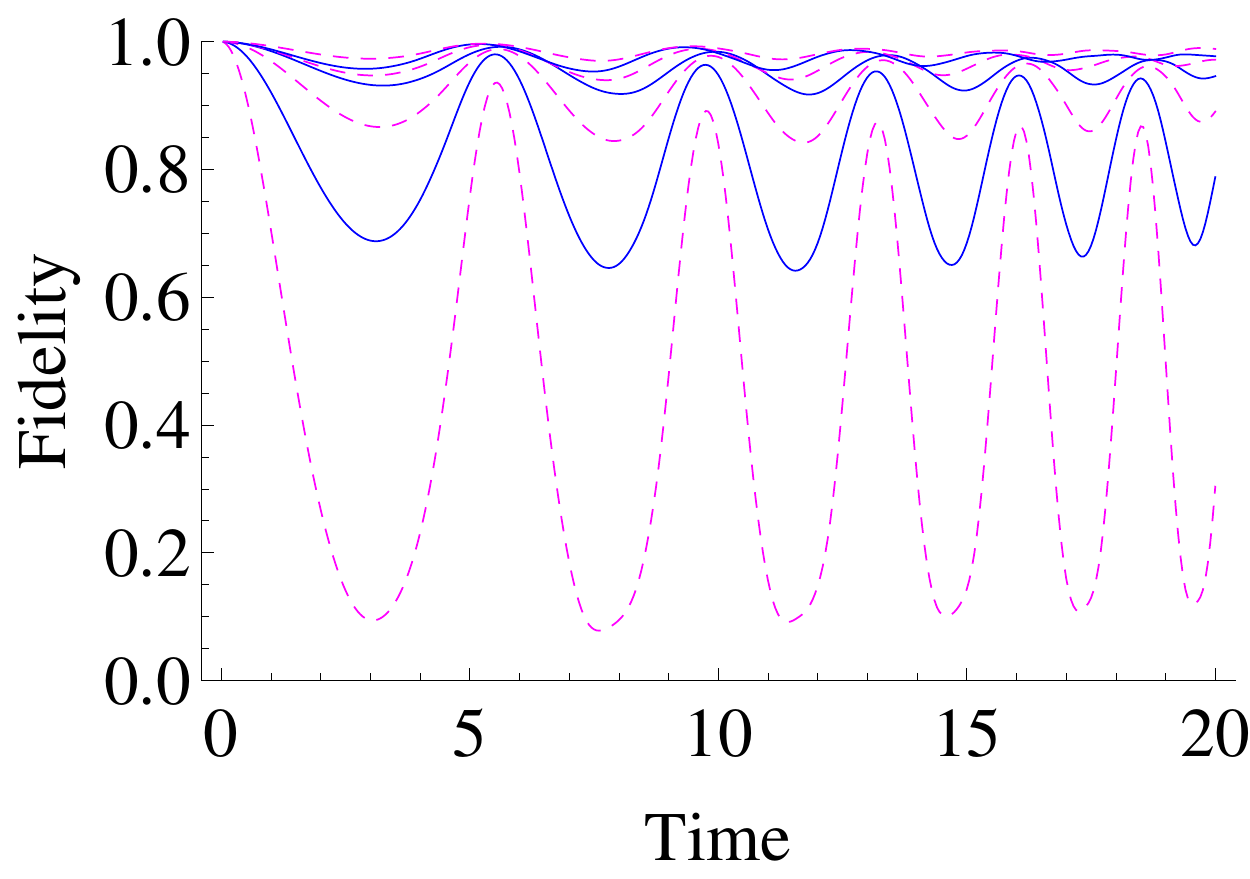} 
\caption{\label{Fig:wocdfid} Evolution of the fidelity under $\hat{H}_0(t)$, for the simulations described in Table \ref{table}. The lowermost (dashed magenta) curve corresponds to $\hbar=1.0$, the next one up (solid blue) corresponds to $\hbar=2.0$, and so forth up to $\hbar=7.0$, which is the magenta curve that remains closest to unity.}
\end{figure}

Fig.\ref{Fig:wocdfid} shows ${\cal F}(t)$ for the seven simulations of evolution under $\hat H_0(t)$ that are represented in the rightmost column of Table \ref{table}.
These plots confirm that the fidelity improves with increasing $\hbar$ (in the absence of the counterdiabatic term), and they display similar oscillatory behavior, with peaks and valleys occurring at nearly the same times for the seven curves.
These observations are consistent with the interpretation that the frequency of the oscillations reflect the corresponding classical evolution, while the magnitude is governed by the quantum energy spacing.

\begin{table}
\begin{center}
\begin{tabular}{|c|c|c|c|c|c|}
\hline
$\hbar$ & $n$ & $\mathcal{F}^{wcd}_{min}$ & $\mathcal{F}^{wocd}_{min}$  \\ \hline
1.0 & 70  & 0.999 & 0.092 \\
2.0 & 35  & 0.999 & 0.641 \\
3.0 & 23  & 0.999 & 0.842 \\
4.0 & 17  & 0.997 & 0.917 \\
5.0 & 14  & 0.992 & 0.939 \\
6.0 & 12  & 0.979 & 0.953 \\
7.0 & 10  & 0.943 & 0.970 \\
\hline 
\end{tabular}
\end{center}
\caption{The dependence of fidelity on the value of the reduced Planck's constant $\hbar$, keeping classical parameters fixed.
The initial quantum number $n$ is chosen such that the initial energy is $E_n \approx 80$. Each simulation is performed at fixed $s=3.0$, while the box length is
varied from $L=25.0$ to $L=15.0$ at $\dot{L}=-0.5$.  $\mathcal{F}^{wcd}_{min}$ is the minimum fidelity when the system evolves under 
$\hat{\mathcal{H}}(t)$, and $\mathcal{F}^{wocd}_{min}$ is the minimum fidelity when the system evolves under $\hat{H}_0(t)$.}
\label{table}
\end{table}

\section{Conclusion}

We have studied a model system undergoing non-scale-invariant driving: the one-dimensional tilted piston described by the Hamiltonian $H_0(q,p;{\boldsymbol\lambda}(t))$ (Eq.(\ref{Hamiltonian})).
We derived exact, closed-form expressions for the classical counter-diabatic Hamiltonian $H_{CD}(q,p,t)$, which we quantized to obtain a Hermitian operator $\hat H_{CD}(t)$. 
In numerical simulations of the time-dependent Schr\"odinger equation, we compared evolution under $\hat H_0$ to that under $\hat{\cal H} = \hat H_0 + \hat H_{CD}$, with the system initialized in an energy eigenstate.
The simulations reveal that $\hat H_{CD}$ very effectively suppresses non-adiabatic transitions: when evolving under $\hat{\cal H}$, the system remains in an eigenstate of $\hat H_0$ with nearly perfect fidelity.
These results establish a proof of principle -- beyond the regime of scale-invariant driving \cite{dCampo13,Deffner14} -- that quantum counterdiabatic Hamiltonians can successfully be constructed from their classical counterparts.

For most Hamiltonians $H_0(z;{\boldsymbol\lambda}(t))$ of interest, a closed-form expression for $H_{CD}(z,t)$ will not be available.
Even for the quite simple system we have studied, the expression for $H_{CD}$ is somewhat involved, and the final result for the operator $\hat{H}_{CD} = \dot{\boldsymbol\lambda}\cdot\hat{\xi}_{SC}$ -- while given in terms of $\hat q$ and $\hat p$ (Eq.(\ref{eq:scHCD})) rather than as a spectral sum (Eq.(\ref{Hqlambda})) -- would certainly be difficult to implement in a laboratory setting.

The difficulty in realizing $\hat{H}_{CD}$ experimentally arises not just because it is given by a complicated expression, but because it is {\it non-local}, i.e.\ because it depends on the momentum operator, $\hat p$.
This is also true for the much simpler counterdiabatic Hamiltonians that have been derived for scale-invariant driving (Eq.(\ref{HCD_scaleInvt})), including the harmonic oscillator (Eq.(\ref{SHOpq})) as a particular example.
In the scale-invariant case, an appropriate canonical (or unitary) transformation of $H_{CD}(q,p,t)$ gives a {\it local} counterdiabatic potential $U_{CD}(q,t)$; in effect, the non-locality can be ``gauged away'' to construct a local shortcut to adiabaticity, as described in Ref.~\cite{Deffner14}.
Whether a transformation of this sort could be applied to our counterdiabatic Hamiltonian $\hat{H}_{CD}$ (Eq.(\ref{eq:scHCD})) is an open question.

\section*{Acknowledgments}

We acknowledge financial support from the U.S. National Science Foundation under grant DMR-1506969 (CJ), and the U.S. Army Research Office under contract number W911NF-13-1-0390 (AP).

\section*{Appendix - Sine basis representation of $\hat{\eta}$}

Because the classical function $\eta(q,p)=\textrm{sign}(p)$ is non-analytic, the matrix representation of its quantal counterpart $\hat\eta$ cannot be obtained by a procedure
like the one used in Sec.~\ref{sec:solving}.
Here we instead construct the matrix representation of $\eta$ by equating its classical and quantum auto-correlation functions.  

Consider a quantum particle in a box with a flat base ($s=0$) and hard walls at $q=0$ and $q=L$, described by the Hamiltonian 
$\hat{H}'=\hat{p}^2/2m + \Theta(\hat{q};0,L)$.
Following Ref.~\cite{Feingold1986}, we write the quantum auto-correlation function  of $\hat{\eta}$, for the eigenstate $|\alpha \rangle$, as
\ba
C_{\alpha}(\tau) 
 &=& \bra \alpha|\hat{\eta} \exp \left(\frac{i \hat{H}' \tau}{\hbar} \right) \hat{\eta} \exp \left(-\frac{i \hat{H}' \tau}{\hbar} \right)|\alpha \ket \non \\
 &=& \sum\limits_{\beta} |\tilde{\eta}_{\alpha \beta}|^2 \exp \left[\frac{i (E_{\beta}-E_{\alpha})\tau}{\hbar} \right],
\ea
where $\tilde{\eta}_{\alpha \beta} = \bra \alpha| \hat{\eta}|\beta \ket$, and $E_{\alpha}$ is the energy corresponding to the eigenstate $|\alpha \rangle$.
The Fourier transform of the auto-correlation function is
\be
\mathbb{C}_{\alpha}(\omega) = \sum\limits_{\beta} |\tilde{\eta}_{\alpha \beta}|^2 \delta(\omega - \omega_{\alpha \beta}),
\label{FT}
\ee
where 
\be 
\omega_{\alpha \beta}\equiv\frac{E_{\beta}-E_{\alpha}}{\hbar}.
\label{omega_ab}
\ee

\begin{figure} 
\centering
\includegraphics[width=0.7\textwidth]{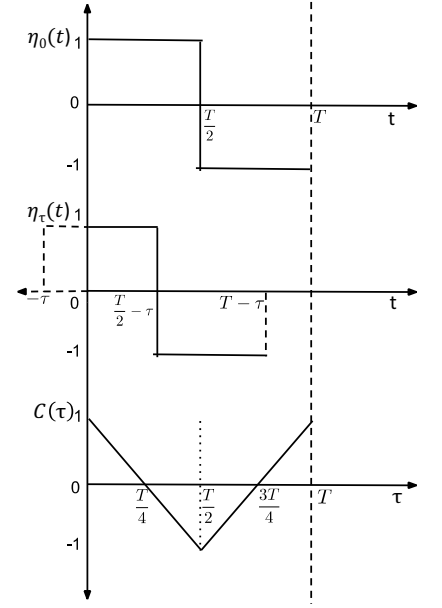} 
\caption{\label{Fig:ACF}The function $\eta_0(t)$ plotted over one time period of oscillation is a square wave (top figure). 
The function $\eta_{\tau}(t)$ is obtained by shifting this square wave leftward by an amount $\tau$ (middle figure).
The autocorrelation function $C(\tau)$ is the product of these square wave pulses, integrated over one period, yielding a triangular wave (bottom figure).} 
\end{figure} 

For a classical particle evolving under the equivalent Hamiltonian, $\eta = \textrm{sign}(p)$ is a square wave pulse with unit amplitude over 
a time period around the energy shell. The functions $\eta_0^E(t)$ and $\eta_{\tau}^E(t)$ describe the dependence of $\eta$ on time for a particle of energy $E$ that 
starts from $L=0$ at times $t= 0$ and  $t=- \tau $ respectively, as depicted in Fig.\ref{Fig:ACF}. 
The classical auto-correlation function, $C_E(\tau) = (1/T) \int_0^{T} dt \, \eta_0^E (t) \eta_{\tau}^E(t)$, is a triangular wave given by
\ba
C_E(\tau)=
\left\{
\begin{array}{lr}
\frac{T-4 \tau}{T}  & ,\quad 0 \le \tau \le \frac{T}{2}\\
\frac{4 \tau-3T}{T}  & ,\quad \frac{T}{2} \le \tau \le T\\
\end{array} ,
\right.
\ea
shown in Fig.\ref{Fig:ACF}.
The Fourier transform of $C_E(\tau)$ is 
\be
\mathbb{C}_E(\omega)=  \sum_{\textrm{odd}\ \gamma=- \infty}^{\infty} \frac{4}{ \pi ^2 \gamma^2} \delta(\omega-\omega_\gamma),
\label{ftc}
\ee
where 
\be 
\omega_{\gamma}=\frac{2 \pi \gamma}{T}.
\label{omega_gamma}
\ee

The correspondence principle suggests that the functions $\mathbb{C}_{\alpha}(\omega)$ and $\mathbb{C}_E(\omega)$ ought to be equal, in the semiclassical limit, when $E_\alpha = E$.
To compare these functions, we first note that for one dimensional systems, the classical action $J(E) = \oint_E p\cdot dq$ satisfies 
\be 
\frac{dJ}{dE}=T.
\label{djde}
\ee
For neighboring energy levels $|\alpha \rangle$ and  $|\alpha +1\rangle$, the energy spacing is
\be
dE = E_{\alpha+1}-E_{\alpha} = \hbar \omega_{\alpha,\alpha+1},
\label{de}
\ee
and the action spacing is given by the Bohr-Sommerfeld quantization condition:
\be
dJ = 2 \pi \hbar.
\label{dj}
\ee
From Eqs.(\ref{djde}) - (\ref{dj}) we obtain $\omega_{\alpha,\alpha+1} = 2 \pi/T$, which generalizes to
\be
\omega_{\alpha\beta} = \frac{2 \pi(\beta - \alpha)}{T},
\label{omega_alphabeta}
\ee
provided $\alpha$ and $\beta$ are not too far apart.

Comparing Eqs.(\ref{omega_gamma}) and (\ref{omega_alphabeta}) we confirm that the delta-functions in Eqs.(\ref{FT}) and (\ref{ftc}) appear at the same frequencies,
and by equating the coefficients of these delta-functions we obtain
\ba
|\tilde{\eta}_{\alpha \beta}| =
\left\{
\begin{array}{lr}
\frac{2}{|\alpha-\beta| \pi} & \alpha -\beta = odd\\
0 & \alpha -\beta = even\\
\end{array}.
\right.
\ea
To ensure that the operator $\hat{\eta}$ is Hermitian (as it represents a physical observable), we impose the condition
$\tilde{\eta}_{\alpha \beta}= \tilde{\eta}_{\beta \alpha}^*$, which then implies  
\ba
\label{etaplusminus}
\tilde{\eta}_{\alpha \beta} =
\left\{
\begin{array}{lr}
\pm \frac{2 i}{(\alpha - \beta) \pi} & \alpha -\beta = odd\\
0 & \alpha -\beta = even\\
\end{array}
\right.
\ea
Finally to determine the sign in Eq.(\ref{etaplusminus}), the ground state eigenfunction of $\hat{H}'(t)$ was boosted by a momentum $p= \pi k/L$, where $k \in 
\mathbb{Z}$, which results in the wave packet $\psi(q)=\sqrt{\frac{2}{L}} \sin(\frac{\pi q}{L}) \exp(\frac{i \pi k q}{L})$. By demanding that 
$\bra \psi|\hat{\eta}| \psi \ket \rightarrow 1 $ for $k \gg 1$  and $\bra \psi|\hat{\eta}| \psi \ket \rightarrow -1 $ for $k \ll -1$, a series of straightforward 
calculations yields
\ba
\tilde{\eta}_{\alpha \beta} =
\left\{
\begin{array}{lr}
\frac{2 i}{(\beta-\alpha) \pi} & \alpha -\beta = odd\\
0 & \alpha -\beta = even\\
\end{array}
\right.
\ea

\end{document}